# Compact Model of Nanowire Tunneling FETs Including Phonon-Assisted Tunneling and Quantum Capacitance

Qiming Shao, Can Zhao, Jinyu Zhang*, Li Zhang, and Zhiping Yu

Institute of Microelectronics, Tsinghua University, Beijing 100084, China

Tel: +86-10-6277-1283, Fax: +86-10-6279-5104, Email: zhangjinyu@tsinghua.edu.cn

**Abstract**

A physics-based compact model for silicon gate-all-around (GAA) nanowire tunneling FETs (NW-tFETs) with good accuracy has been developed by considering Phonon-Assisted Tunneling (PAT) and transition from Quantum Capacitance Limit (QCL) to Classical Limit (CL) during the device-size scaling. The impact of PAT results in the broadening of a single electron-energy level to an energy band with density-of-states (DOS) distribution of Lorentzian shape. As a consequence, the tunneling probability at the edge of tunneling window no longer changes abruptly from zero to having a finite value. By adjusting the parameters in the Lorentzian function, an accurate fitting to the measured transfer characteristics in the subthreshold region is made possible. Besides, with an analytical formula to calculate the channel potential, the model is able to cover naturally the transition from QCL to CL regime when the device size is scaled. Furthermore, on-voltage is defined to facilitate the modeling and fitting processes. Comparisons with the experimental data demonstrate the model accuracy across all device operation regions and the flexibility in model parameter extraction is also shown.

# I. Introduction

The GAA NW-MOSFETs have been widely investigated to reduce the short-channel effects [1], yet the minimum subthreshold swing (SS) is still limited to 60mV/dec due to the thermionic-emission nature of pn junctions. To continue the scaling trend, tunneling-FETs (tFETs) have been proposed to overcome this SS limit [2]. Compact models for double-gate (DG) and NW-tFETs have been published in [3] and [4], but the comparisons to experiments lack. While our previous model [5] achieved qualitative match with experimental data [6], the quantitative fitting to the non-ideal effects is not satisfactory. In this article, we propose a new and more accurate model by including the phonon-assisted tunneling (PAT) mechanism and the model covers the transition from the classic limit (CL) to the quantum capacitance limit (QCL) as the device size is scaled down. The accuracy of the modeling results is vastly improved in comparison with the measured data on all operation regions of NW-tFETs. Moreover, a simple extraction scheme for model parameters is also developed to make this model flexible enough for practical use. In Section II, we review the basic compact model without PAT and quantum capacitance as developed in [5], in which definition of on-voltage is first introduced. In Sections III and IV, we describe how PAT and quantum capacitance can be included in the compact model to improve the model accuracy. In Section V, we compare the modeling results with experimental data in [6-7] to verify this compact model. Besides, how fitting parameters affect the performance of NW-tFETs beyond the available experimental data is discussed, and the conclusion is made in Section VI.

# II. Compact Model Description

In Figs. 1(a-b), schematics of n/p-type of NW-tFETs are shown. The source and drain regions are doped degenerately, not only to keep device contact resistances small, but also to make the tunneling through the source/channel junction efficient. In this paper, only n-type of NW-tFETs (and thus the doping in the source region is $p^+$) are discussed for model development. The band diagrams along the channel in OFF and ON states are shown in Figs. 2(a-b). The differentiation of two states lies on the relative position of the top of valence band in the source

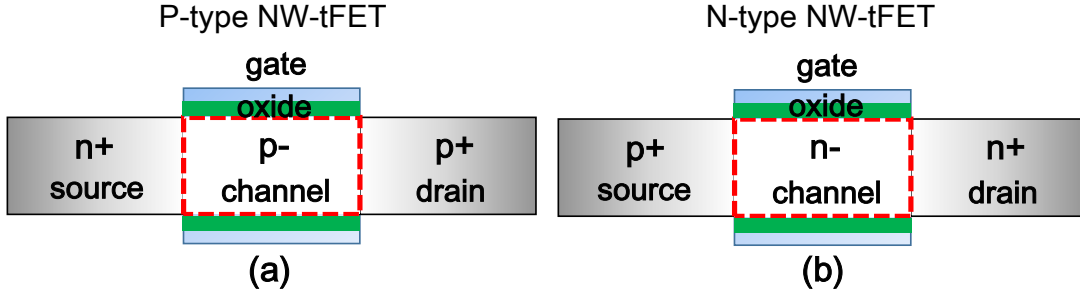

Fig.1 Sectional views of NW-tFETs. (a) n-type (b) p-type

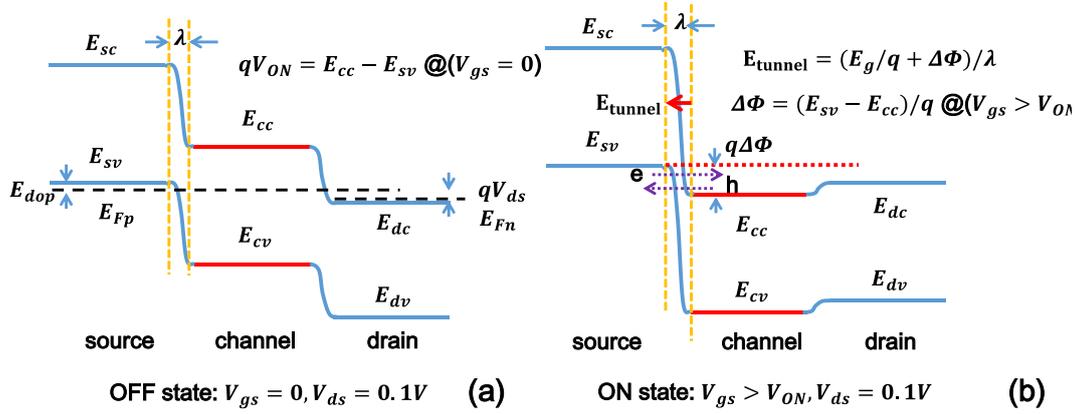

Fig.2 Band diagram along the channel for n-type NW-tFETs. (a) OFF state (b) ON state

region ($E_{sv}$) with respect to the bottom of the conduction band in the channel ($E_{cc}$): $E_{sv}-E_{cc}>0$ for ON state and $E_{sv}-E_{cc}<0$ for OFF state. To further distinguish the OFF and ON states, we define an on-voltage ($V_{ON}$) for gate voltage $V_{gs}$ [5], when $E_{sv}=E_{cc}$. The parameter $V_{ON}$ signifies the sudden rise of the drain current in a semi-logarithmic transfer plot of $I_{ds}$ vs. $V_{gs}$ (refer to Fig. 5). This value depends on $E_g$ (the bandgap in the channel region), $\Delta\varphi$ (difference of work functions between the gate electrode and channel), $E_{ini\_ch}$ (doping level in the channel), $E_{dop}$ (difference between the Fermi level and $E_{sv}$ in the source region, see Fig. 2a); $qV_{ON} = E_g/2 - E_{dop} - E_{ini\_ch} + \Delta\varphi$. All these parameters are chosen such that when the $V_{gs}$ is zero, $E_{cc}$ is above $E_{sv}$, and thus the device is on OFF state (no tunneling current). Nonetheless, there is still some leakage current measured, which we take as the model parameter $I_{off}$. Besides above factors, $V_{ON}$ can be affected by drain-source bias ($V_{ds}$), which is similar with the DIBL effect and can be observed in the measured transfer curves. We then have $E_{cc}(V_{gs}) = q(V_{ON}-k_{DIBL}V_{ds})-q\phi_s(V_{gs})$,

with $k_{DIBL}$ a model parameter and $\phi_s$ the channel potential. Here, $\phi_s$ is treated as a variable depending only on the gate voltage. In three-dimensional (3D) case, however, $\phi_s$ is the channel surface potential and varies along the channel. The exact relation between $V_{gs}$ and $\phi_s$ is considered in Section IV. As $V_{gs}$ is increased, $E_{cc}$ is pressed down, and when $E_{cc}$ becomes lying below $E_{sv}$, a tunneling window $\Delta\Phi$ (defined in Fig. 1b) is opened and there will be a tunneling current for non-zero $V_{ds}$.

The drain current is determined by Landauer's formula [8] for ballistic transport without PAT,

$$I = \frac{2q}{h} \int_{E_{cc}}^{E_{sv}} T_S T_{BTBT} [f_D(E) - f_S(E)] dE, \qquad (1)$$

where $h$ is the Planck constant, product $T_S T_{BTBT}$ denotes the total transmission probability (independent upon the tunneling energy) with $T_S$ the diffusive coefficient to account for the ballisticity and $T_S = T_{scat}/(L_{scat}+L)$, where $L_{scat}$ is the mean free path for carriers in silicon nanowire (assuming to be about 110nm in our case [9]). $T_{BTBT}$ is calculated using WKB approximation and the tunneling barrier is assumed to be triagonal [10],

$$T_{BTBT} = e^{-4\lambda \sqrt{2m_r E_g^3}/3\hbar(E_g + q\Delta\Phi)}, \qquad (2)$$

where $m_r$ is the relative effective mass for carriers in the barrier, and $\lambda$ is a critical parameter, which appears as the length scaling factor. In GAA device structure, $\lambda = \sqrt{[2\varepsilon_{Si} d_{Si}^2 \ln(1 + 2t_{ox}/d_{Si}) + \varepsilon_{ox} d_{Si}^2]/16\varepsilon_{ox}}$, in which $t_{ox}$ is the thickness of the gate oxide, $d_{si}$ is the diameter of nanowire, $\varepsilon_{Si}$ and $\varepsilon_{ox}$ are the dielectric constants of silicon and oxide, respectively [11]. $f_{D/S}(E)$ represent the Fermi-Dirac functions with reference to Fermi-levels at drain and source regions, respectively.

### III. Model for Phonon-Assisted Tunneling

In the previous section, a quantity of $V_{ON}$ has been defined, and its significance is that this voltage indicates at which gate voltage the tunneling window starts to open. However, calculation using the conventional Zener tunneling formula [10] predicts too steep a rise in subthreshold region compared to measurement (see Fig. 6, boxed area). We tracked this discrepancy to the failure in incorporating the phonon-assisted tunneling (PAT) in the original theory. To make the remedy in a concise way, we have developed an analytical approach to incorporating the PAT effect:

$$I_{ds} = \frac{2q}{h} T_S T_{BTBT} \int_{-\infty}^{E_{sv}} [f_S(E) - f_D(E)] dE \int_{E_{cc}}^{+\infty} \frac{1}{\pi} \frac{\Gamma/2}{(E-E_0)^2 + (\Gamma/2)^2}, \quad (3)$$

i.e., in the integrand for the integral in $E$, we have added a multiplication factor (the second integral), which itself is an integral with $E$ as the parameter. In the above formula, $\Gamma$ is the effective broadening width of the Lorentzian function, which is related with carrier scattering rate [12], and thus dependent upon the diameter of nanowire [9]. Thanks to the nature of Lorentzian function, this multiplication factor could be analytically computed as $1/2 - 1/\pi *\arctan[2(E_{cc}-E)/\Gamma]$. The magnitude of $\Gamma$ denotes the PAT effect on the device performance. The reason why PAT helps slow the current rise in the subthreshold region can be explained as follows: without PAT, the tunneling window opens abruptly for if there is no corresponding energy state on the receiving side to which the carrier is to tunnel, the tunneling probability is strictly zero (Fig. 4a). With PAT, however, the energy for carriers on the launch side is no longer limited to a single value, rather the carrier energy is smeared out to become a band of finite width (Fig. 3). We propose to use Lorentzian distribution for the shape of this energy band (Fig. 4b), since it physically describes collision broadening quite well [12]. The only model parameter is the broadening factor $\Gamma$ in the Lorentzian function (Fig. 4b), which is used as a fitting parameter. If $\Gamma$ equals zero, the multiplication factor is a step function; when $E>E_{cc}$, it is unity, and zero otherwise. This is the picture without PAT. If $\Gamma$ is nonzero, the multiplication factor will smoothly increase when $E$ increase towards $E_{cc}$.

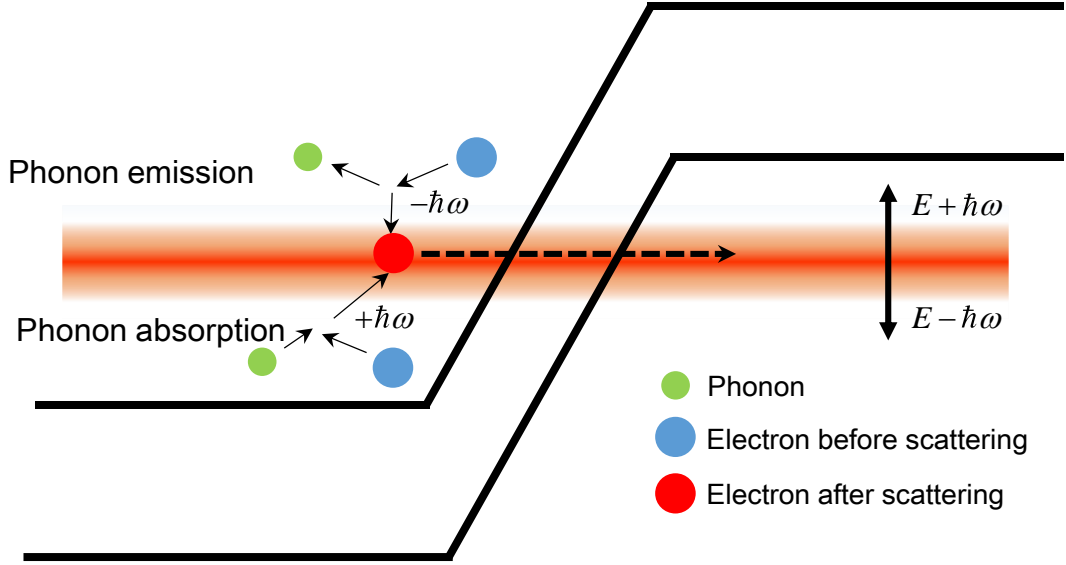

Fig.3 Schematic of phonon-assisted tunneling (PAT). During PAT process, total energy and momentum conservation of phonon and electron are observed.

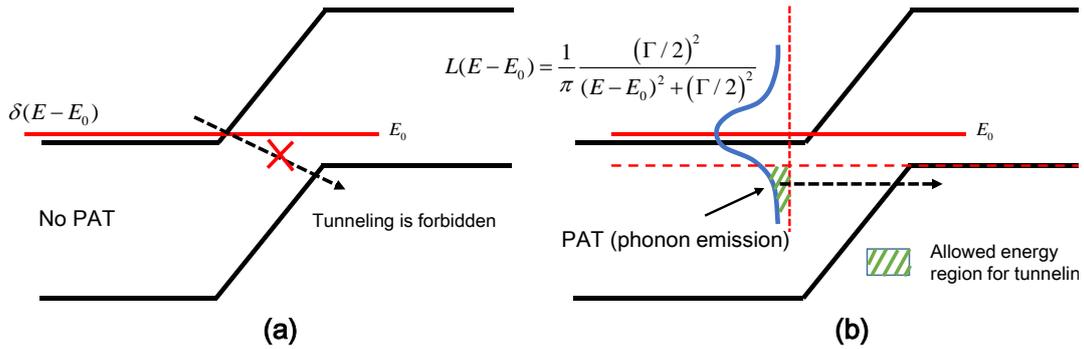

Fig.4 Comparison of model with and without PAT. (a) Without PAT, there is no tunneling outside of the tunneling window. (b) The process of PAT is modeled as the broadening of electron energy from a single level to a distributed form with Lorentzian shape. Even when there is no tunneling window, with PAT, tunneling can still occur.

### IV. Model Scalability: from Classic Limit to Quantum Capacitance Limit

The gate-voltage controlled channel potential, $\phi_s$, is a critical factor in our compact model. In general, the total charge in the channel $Q_{tot}$ is given by $Q_{tot} = \phi_s(C_s+C_{ox}+C_d) = C_sV_s+C_dV_d+C_{ox}V_g+Q_{ch}$, in which $C_s$, $C_d$, $C_{ox}$ and $Q_{ch}$ are source capacitance, drain capacitance, oxide capacitance and mobile charge, respectively [13]. Note that for GAA gate capacitance, $C_{ox} = \varepsilon_{ox}/[d_{Si}/2*\ln(1+2t_{ox}/d_{Si})]$. Quantum capacitance $C_q$ in our model is given by $C_q = -\partial Q_{ch}/\partial \phi_s$, which is proportional to the density of states (DOS) [14]. To preserve the model accuracy while the device is scaled down from the classical regime to quantum confinement regime, we developed an analytical formula for the rate of change in $\phi_s$ with respect to $V_{gs}$, i.e., $d\phi_s/dV_{gs}$, and use a single fitting parameter $\alpha$ to express its dependence on $V_{ds}$ and $\phi_s$:

$$\frac{d\phi_s}{dV_{gs}} \approx \frac{1}{1+C_q/C_{ox}} \cong \frac{1}{1+\frac{e^2}{2\pi}\left(\frac{2m_r}{\hbar^2}\right)^{1+\alpha}\left(\frac{d_{Si}}{2}\right)^{2\alpha}\phi_s^\alpha/C_{ox}}, \qquad (4)$$

Where $\alpha$ stands for the quantum capacitance index. Note that assumption $C_{s,d} \ll C_{ox}$ have been used, which is normally the case in electrostatically well-behaved devices [13]. In 1D case, $\alpha = -1/2$ and in 3D case, $\alpha = 1/2$ (the rational: the DOS of 1D system is proportional to $E^{-1/2}$, and the DOS of 3D system is proportional to $E^{1/2}$). In two extreme cases, $\alpha=-0.5$, $d\phi_s/dV_{gs} \to 1$ for ideal QCL, and $\alpha=0.5$, $d\phi_s/dV_{gs} \to 0$ for a complete 3D structure. By solving this nonlinear differential equation (4) (the boundary condition is trivial: when $V_{gs}=V_{ON}$, $\phi_s = V_{ON}$), now we can obtain the channel potential $\phi_s$.

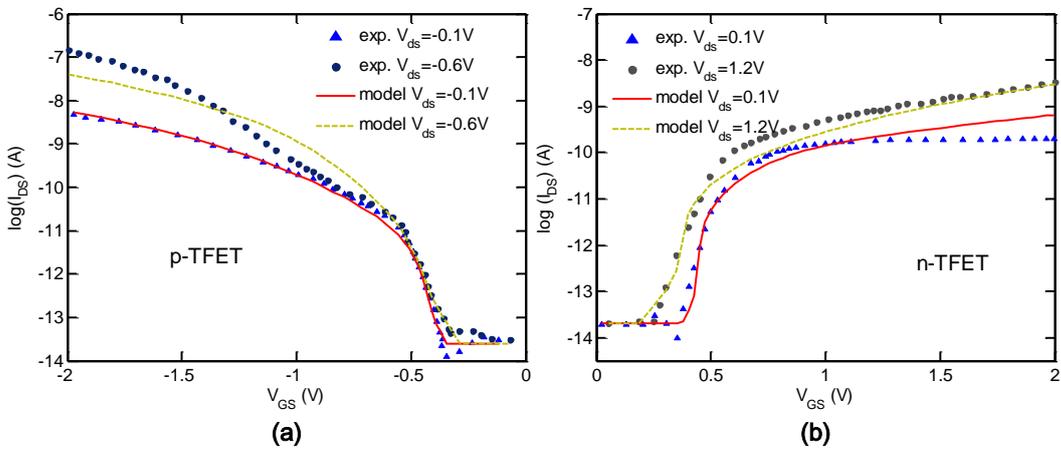

Fig. 5. Comparison between compact model and experimental data [2-3]. (a) p-type and (b) n-type NW-tFETs. It can be seen that experimental p-type and n-type NW-tFETs operate as in QCL and CL regimes, respectively, which is consistent with the fact that the diameter of nanowire in p-type device is 18nm, much smaller than the size of nanowire in n-type, 40nm.

## V. Model Verification and Discussion

In [6-7], vertical p/n NW-tFETs are fabricated and the measured transfer I-V characteristics are available. The results from our model evaluation (including parameter extraction for $V_{ON}$ and $I_{off}$) are compared to these data. Good agreement has been achieved for most cases, as shown in Fig. 5. In table I, the fitting parameters are given. To further explain how parameters $\Gamma$, $k_{DIBL}$ and $\alpha$ affect the accuracy, we consider their impact on the model evaluation separately. Shown in Fig. 6 is the comparison of calculated results from the model with and without considering PAT. Without PAT, the drain current rises steeply when the gate voltage is just passed the on-voltage, since there is an abrupt change of tunneling possibility from zero to a finite value. After considering the PAT, the current increases relatively slowly, which is more close to what the experiment reveals. The theoretical analysis by NEGF simulation [15] also confirms this gentle rise. Fig. 7 demonstrates the influence of parameter $k_{DIBL}$ on the transfer characteristics for different drain voltage. The on-voltage decreases as the drain voltage increases, especially with larger $k_{DIBL}$. Fig. 8 describes the transition from QCL to CL during the device-size scaling. When the nanowire is scaled to a small size (in diameter), the impact of the gate voltage on the channel potential becomes large. Even when Fermi level enters into the valence band, the gate voltage can still effectively tune the tunneling window in a finite range, and thus better performance is achieved: the lower the on-voltage, the steeper the SS, and the higher the on-current (see Fig. 8). The demarcation between QCL ($k_{DIBL}=0$, $\alpha=-0.5$) and CL ($k_{DIBL}\neq0$, $\alpha=0.5$) is largely empirical for the model at the present modeling state.

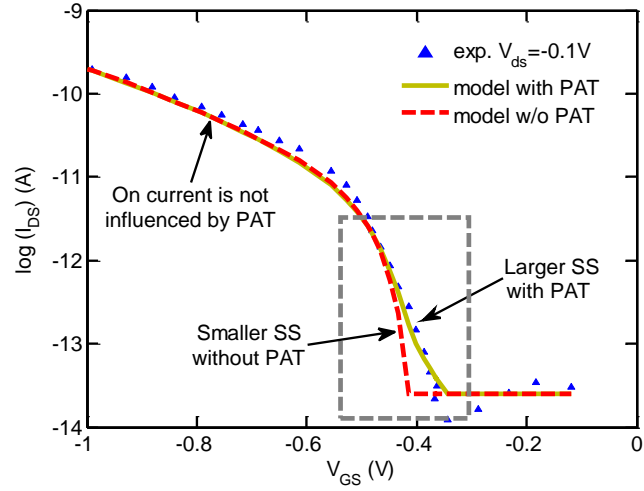

Fig.6 Fitting results with and without of PAT, which shows PAT helps improve accuracy of model. For PAT, Γ=20meV.

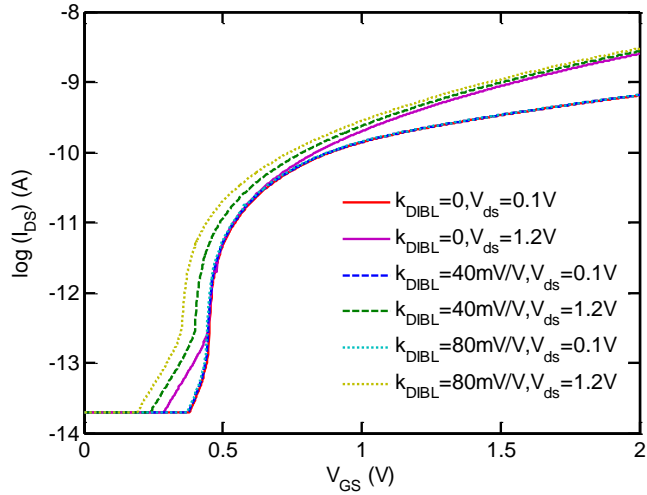

Fig.7 Dependence of device transfer characteristics on $k_{DIBL}$. $k_{DIBL}$ is a parameter in describing DIBL effect where larger $k_{DIBL}$ gives stronger DIBL effect.

**Table I.** Parameters for matched n/p NW-tFETs in Fig. 5

| Parameters | p-tFET | n-tFET |
| --- | --- | --- |
| $I_{off}$ (A) | $2.5*10^{-14}$ | $2*10^{-14}$ |
| $m_r$ ($m_{e0}$) | 0.053 | 0.01 |
| $V_{ON}$(V) | -0.42 | 0.45 |
| Γ (meV) | 20 | 0.6 |
| $k_{DIBL}$ (mV/V) | 0 | 80 |
| α | -0.07 | 0.033 |

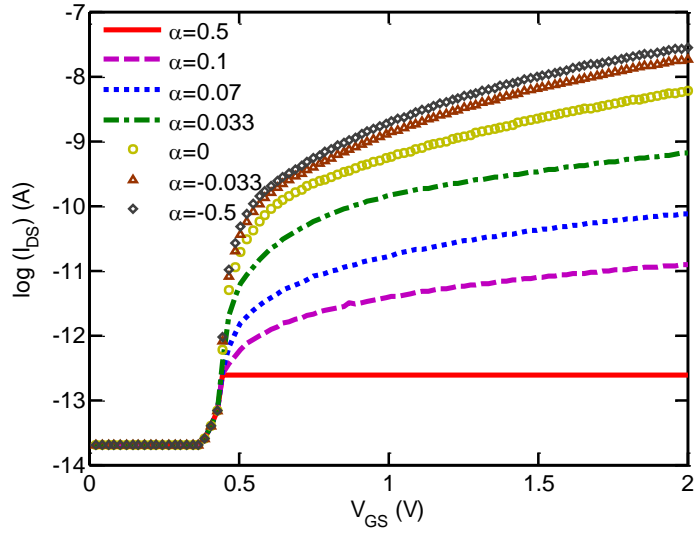

Fig.8 Dependence of device transfer characteristics on $\alpha$. $\alpha$ denotes the transition from QCL to CL. Larger $\alpha$ means the device operates more in CL regime. QCL is outperformed than CL when $\alpha$ approaches to -0.5.

Nonetheless, the introduction of parameters $k_{\text{DIBL}}$ and $\alpha$ is enough to cover the transition between QCL and CL.

**VI. Conclusion**

In this article, a complete compact model of NW-tFETs with good accuracy in fitting experimental data is proposed, and the model incorporates the impact of PAT and scaling on the device performance. In addition, the definition of on-voltage helps understand the concept of another type of "threshold voltage" in NW-tFETs and facilitates the fitting to the experiment. This model can serve the base for a full-blown and practical compact model for GAA-NW tFETs.

**Acknowledge**

This work is supported by China's "National Key Fundamental Research Development Project (973 Project)" #2011CBA00604.